\documentclass{article} 

 \usepackage{graphicx}
 \usepackage{setspace}   

\usepackage{amsmath} 
\usepackage{amssymb} 
\usepackage{amsthm} 
\usepackage{amscd} 

\author{Edoh Y.~Amiran\footnote{Department of Mathematics, Western Washington University, 
             516 High Street,
             Bellingham, WA, 98225-9063
             USA, edoh@wwu.edu } \ and Joni S.~James Charles\footnote{ jc18@txstate.edu} }

\begin{document}

\title{\large Reconciling revealed and stated measures for willingness to pay in recreation by building a probability model.}


\maketitle

\bigskip \noindent 
{\bf Abstract}:
The consumers' willingness to pay plays an important role in economic theory and in setting policy.  For a market, this function can often be estimated from observed behavior -- preferences are revealed. However, economists would like to measure consumers' willingness to pay for some goods where this can only be measured through stated valuation. Confirmed convergence of valuations based on stated preferences as compared to valuations based on revealed preferences is rare, and it is important to establish circumstances under which one can expect such convergence.  By building a simple probabilistic model for the consumers' likelihood of travel, we provide an approach that should make comparing stated and revealed preferences easier in cases where the preference is tied to travel or some other behavior whose cost can be measured.  We implemented this approach in a pilot study and found an estimate of willingness to pay for visiting an environmentally enhanced recreational site based on actual travel in good agreement with an estimate based on a survey using stated preferences.  To use the probabilistic model we used population statistics to adjust for the relevant duration and thus compare stated and revealed responses. 





\section{Revealed and Stated Preferences.}

In cost-benefit analysis and for valuations that inform policy, the relevant quantity is the change in the amount of the provided good and the corresponding expenditure. One can (in theory) measure the maximum that consumers would be willing to pay for a given additional amount of the good, which is abbreviated as WTP, and this is the economic benefit to consumers. More precisely, when a certain quantity of the good is available, and a particular increment in the good is provided, and one can measure the maximal amount that consumers will pay for that increment, then we have a measure for WTP. 

\bigskip \noindent 
For example, suppose that a recreational site is available and consumers visit the site, incurring costs for visiting, such as the cost of travel, the time it takes to travel, expenditures on food and lodging for the visit, needed equipment, and so on.  Now suppose that gas prices rise thereby increasing the travel cost, but no other factor changes.  Then we learn how many of the visitors continue to visit, accepting the additional cost of the additional visit, and how many do not.  This additional cost is associated with an additional visit, besides the ones previously undertaken.  Hence, potentially, we could find WTP for an additional visit (assuming that we know the base pattern of visits, before the rise in travel cost). 

\bigskip \noindent 
An alternative to finding WTP for a particular increment from particular starting quantities is to find the entire demand curve for the good of interest.  This, of course, requires data that is often difficult to obtain. When demand can be modeled and the model calibrated using available data, one may obtain sufficient accuracy in a range of values relevant for particular policy decisions. For a market good the more general demand can at times be estimated using variations occurring in different markets, or by varying prices either directly or through coupons, or through auctions. 

\bigskip \noindent 
When WTP is measured from direct observations on consumer behavior when increments are introduced, or from an estimated demand function based on observations, one does not need the added complication of constructing a model of the underlying behavior -- making assumptions about the process through which the consumer decides whether to increase the amount consumed. However, if one can make reasonable assumptions about the underlying process, and the data supports retaining those assumptions, then the calibration requires less data and the model can be applied to a larger range of amounts of the good. 

\bigskip \noindent 
For example, many policy decisions regard investing in additional open space, parks, preserves, or recreational amenities, and it is impossible to measure the benefit directly unless the investment is already made. Hence an underlying model can be useful for policy purposes.  This issue has given rise to the field of benefits transfers, where the benefits estimated directly in one situation are used for estimating the benefits of a proposed project at a similar but different setting, and while this approach avoids the need for a model of behavior, it comes with its own questions and uncertainties. 

\bigskip \noindent 
When estimating WTP, the distinction between increments and the demand may be blurred by the definition of the good.  When the policy decision regards opening a recreational site, is the site the whole good (and, of course, the whole increment)? or are all recreational sites the whole good? and is a recreational site for mountain biking equivalent to one for swimming or boating in a river, or to a bird-watching site.  For instance, if a site dedicated to mountain biking is enhanced we could think of the enhancement as the increment, or we could think of the demand for the particular site as the ``whole''. From the point of view of policy these will provide equally useful estimates, because the increment of interest is included even when the whole of the good is poorly defined.  

\bigskip \noindent 
In summary, WTP is the relevant quantity for policy, and should be measured for a particular increment from a particular starting point (amount of the good and starting expenditure), and yet it can be computed from demand for a larger range of amounts of the good. 

\bigskip \noindent 
Absent an existing market for the good of interest, many elicitation methods used to assess WTP for public goods, and especially for experiential or environmental goods, require a response to hypothetical questions.  The response to a hypothetical question is necessarily a stated preference, meaning a value of WTP based on a statement from the consumer rather than on the consumer's revealed behavior in the market place. 

\bigskip \noindent 
For conclusions from a survey requesting stated preferences to be useful the responses must represent the same construct for the consumers responding and for those interpreting the result, and must represent an identifiable construct -- one that represents an approximation, in the statistical sense, of some explicitly described quality or some theoretically known quantity. 

\bigskip \noindent 
In this paper we describe an approach and a method that allow us to compare the demand for travel to an environmental and experiential site using the observed attendance at the site to an estimate for WTP obtained from a hypothetical question regarding travel under changed conditions.  Therefore, the method allows a comparison between revealed WTP and stated WTP for this one component of valuation of the good. 

\bigskip \noindent 
Our approach uses an underlying model of consumer decisions that gives rise to a probabilistic version of the demand.  In particular, a person located a certain distance from the site providing the amenity is seen as having a probability of visiting the site.  This simple model allows one to both calculate the average (or total) expenditure for people located at that distance and to calculate the response when the cost of travel changes.  Therefore we have the demand for travel to the particular site and a convenient method for computing WTP from it. 

\bigskip \noindent 
By hypothesizing a change in the cost of travel we create a new setting and can ask respondents to state their behavior in this new hypothetical setting.  The response, when interpreted through the model, converts nicely to a statement of the value of travel to the site under an increment in expenditure, and can easily be converted to a measure of WTP. 

\bigskip \noindent 
Our model represents a large departure from the usual approaches to calculating WTP or demand.  Economists have used probabilistic models to analyze demand under uncertainty, but we are aware of no instances where the demand was viewed probabilistically.  The existing models involving uncertainty model the choices made by an individual facing risk, and describe the likelihood of a particular choice, while our model describes the likelihood of consumption. 

\bigskip \noindent 
Our approach does present some difficulties in comparing preferences from a fixed population -- preferences depending on location -- to preferences over consumption for individuals during a fixed period.  These difficulties are resolved without developing new methods, by analyzing the growth of the population over time. 

\bigskip \noindent 
We first explain how the new method works, then discuss the economic literature that gives rise to the original question regarding revealed and stated preferences, illustrate the use of the method with results from a small pilot study, and finally discuss the implications of our findings from this exploratory study. 

\section{The Model for Revealed and Stated WTP from Travel.} 
\label{sec:model} 

Given a sample one can calculate the travel cost for those consumers who have undertaken the travel.  However, the cost paid by these consumers could be either the total amount they are willing to pay (for the quantity consumed) or a smaller amount.  This is typically resolved by estimating the additional amount that consumers are willing to pay for an increment of the good, which requires more than observing the current expenditure on travel. 

\bigskip \noindent 
Moreover, to interpret the result of a hypothetical situation, one needs a prediction for consumers' behavior in the modified setting. 

\bigskip \noindent 
Both of these issues can be addressed by a probabilistic model.  First, under the usual assumption that the population is homogeneous, when a person decides to travel a certain distance, the person provides data on the probability that a (generic) person decides to travel this distance.  The probability is related to the amount that consumers pay, so one can calculate a demand curve. 

\bigskip \noindent 
Second, certain hypothetical situations can be quantified in terms of the parameters in the model and the quantity predicted by the model can be compared to the stated preferences. 

\subsection{The probability model.}
\label{sec:prob} 

Our model assumes that distance traveled to the tourism site is exponentially distributed.  In an appendix we recall the known properties of this model and explain why this commonly used distribution makes sense for travel and travel costs. 

\bigskip \noindent 
Let $X$ denote the total amount spent.  Then WTP is $w$ with probability 
\begin{equation}
{\rm Pr } \big( X \leq w \big) = \int_0^{a\, w} \lambda e^{-\lambda u} \, du \, .
\label{eq:fullExp}
\end{equation}
In this equation, $u$ represents the distance traveled, which has the exponential distribution. 

\bigskip \noindent 
This exponential distribution has mean $1/\lambda$ and the cumulative distribution function $F(u)=1-e^{-\lambda u}$.  The expenditure on transportation is proportional to the distance traveled and the parameter $a$ depends on the units, since when $aX$ is exponentially distributed with mean $1/\lambda$ we have  ${\rm Pr} ( X \leq w ) = {\rm Pr} ( a\, X \leq a\, w )$. 

\bigskip \noindent 
The average, $1/\lambda$, is measured in units of expenditure per tourist.  To estimate $\lambda$, we fit the probabilities in  equation \ref{eq:fullExp}, where the probability of traveling a distance in a certain interval is the number of consumers attending the site as a proportion of the population. In practice, the population is grouped into intervals for the distance, we use the total population whose distance from the site visited falls into an interval, and calculate the probability as the proportion of the population that visits from the corresponding zip codes. 

\bigskip \noindent 
This approach solves the first problem described in the beginning of this section, that is, we calculate the (total) demand for visiting the particular site without requiring an experiment to confirm a maximum expenditure. 

\bigskip \noindent 
In estimating the parameters we have distances, $x_j$ with $1\leq j \leq k$, corresponding to different locations and the corresponding probabilities of visiting, $p_j$. We fit the model $p = \lambda \, e^{-r \, x}$ and obtain $a = r / \, \lambda$ which can be used to estimate the likelihood of a particular expenditure through the corresponding distance of travel.  The revealed WTP obtained in this way represents the demand for an annual visit by a member of the population. 

\subsection{The hypothetical setting.}
\label{sec:change} 

The hypothetical question was, in this example, whether a consumer would continue to visit if fuel costs were to increase.  If a respondent stated that they will continue visiting the site with the higher cost of fuel, then that respondent is willing to pay an additional cost above the one currently paid.  

\bigskip \noindent 
Let $\Delta$ be the additional cost for a traveler whose initial (current) cost is $t$. 

\bigskip \noindent 
The proportion of the population willing to pay $t$ or more for a visit to the tourism site is
\begin{equation}
{\rm P}(t) = \int_{t}^{\infty} \lambda e^{-\lambda x}\, dx = e^{-\lambda t}.
\end{equation}

\bigskip \noindent 
The proportion of the population willing to pay $t+\Delta$ (or more) is $ e^{-\lambda  (t +\Delta )}$ and so the proportion of those willing to pay $t+\Delta$ (or more) to those willing to pay $t$ (or more) is $e^{-\lambda (t+\Delta)} \big/ e^{-\lambda t} = e^{-\lambda \Delta}$.  Notice that this proportion depends only on $\Delta$, and thus is easily found from the probability that a respondent is willing to pay $\Delta$ in additional costs.  In other words, for a particular consumer we need to know the likelihood that this consumer is willing to pay the additional cost, but we do not need to know the current cost paid or that the current cost is the maximum that this consumer is willing to pay. 

\bigskip \noindent 
In our analysis, and necessarily for this methodology, travelers are grouped according to their distance of travel.  Since the additional cost is proportional to the distance of travel, the additional cost varies by group.  Suppose there are $k$ groups and let $\Delta_i$ be the additional cost for the $i^{\rm th}$ group.  As described above, the probability that an individual is willing to pay an additional $\Delta_{i}$ is independent of the amount that the person is already paying.  Denote the probability that a tourist from this group agrees to pay the additional cost by $p_i$.  Denote the number of respondents in the $i^{\rm th}$ group by $n_i$  and the total number of respondents by $\sum_{i}\, n_i = N$. 

\bigskip \noindent 
The average additional cost, $\Delta$, is $\Delta = \frac{1}{N}\sum_{i} n_i\, \Delta_i$ and satisfies 
\begin{equation}
e^{ - \lambda \, \Delta } = e^{ - \lambda \, \frac{1}{N} \sum n_i \, \Delta_i } = \Pi_{i} \, e^{ -(\lambda / N) n_i \,  \Delta _i} = \big(  \Pi_{i} \, e^{ - \lambda n_i \,  \Delta _i}  \big)^{\frac{1}{N}} = \big(  \Pi_{i} \, p_i^{n_i}  \big)^{\frac{1}{N}} .
\label{eq:Delta}
\end{equation} 

\bigskip \noindent 
The numbers $n_i$, the probabilities $p_i$, and $\Delta$ are known from the survey or can be calculated from the responses to the hypothetical question.  Therefore $\lambda$ can be estimates from the stated preferences, separately from the estimated value using revealed preferences. 

\bigskip \noindent 
In the setting of a survey, the amount that a person is willing to pay is the amount for an annual visit from a particular visitor.  In other words, it represents an additional amount per visit, and since we derive an annual amount from the sampled visits, per year, rather than over the lifetime of the visitor.  

\section{Uses of revealed and stated willingness to pay and questions regarding measures.}
\label{sec:measures}

Economists have sought methods for measuring demand for recreation starting no later than \cite{Clawson1959} and continuing with \cite{Knetsch1963} and other studies, such as \cite{ClaK1966} and \cite{Cesario1976}.  

\bigskip \noindent 
Among the measures developed were travel costs which are used to calculate part of the amount that a consumer is willing to pay in order to access a tourism or recreation site.  These are typically estimated at the destination, by asking visitors directly about the costs or by using the distance traveled and the average cost for it.  

\bigskip \noindent 
Other measures developed aimed to capture the value of visitors' experience through other expenditures such as time \cite{Cesario1976}, or to value other aspects of natural or man-made assets.  The measures valuing other aspects typically ask consumers to declare the value they attach to an experience or potential experience or to the existence value associated with a site or resource.  Since the surveys typically present hypothetical settings to the respondents, they are known as contingent valuation methods. 

\bigskip \noindent 
Using more detailed statistical models, experimenters present responders with a variety of choices, including various amounts of environmental and market goods.  These choice experiments aim to more carefully quantify the value that consumers attach to each ingredient.  

\bigskip \noindent 
Contingent valuation and choice experiments play an important role in estimating the value of the existence of natural resources and of other experiential or non-market goods.  Contingent valuation has also been used to estimate the value of aspects of health and longevity.  For examples using environmental goods, and for a discussion of some questions regarding them, see  \cite{HeberleinAnd2005}. For examples regarding health and longevity see \cite{AlberiniChiabai}.  

\bigskip \noindent 
For contingent valuation, responses vary significantly with the description of a setting or the phrasing of questions which reflects discussions in consumer research such as \cite{BettmanKakkar}, and echo earlier observations and questions regarding the relationship of attitudes and behavior such as \cite{LaPiere1934} and \cite{SchumanJohnson1976}. 

\bigskip \noindent 
In particular settings, where combinations of goods may be observed in a ``natural'' experiment, hedonic methods can be used to determine WTP.  For instance, in \cite{ZhangZheng2015} WTP for a non-market social good, diversity, is estimated using variations in choices when consumers move to a new location, and \cite{Black2018} estimates the value of open space, having both use and existence values, using changes in property values when the state (via a process that is difficult to predict) purchases adjacent land. The value of clean air is similarly examined in \cite{NishitatenoBurke} when property values changed with regulations for air quality. 

\bigskip \noindent 
Since contingent valuation had significant implications for policy regarding natural resources, much of the discussion occurred in the environmental economics literature.  An early discussion of surveys and theory is \\ \cite{BishopHeberlein1986}, which was followed by a definitive work on the theory and substance of the method in \cite{MitchellCarson1989}. 

\bigskip \noindent 
Many experiments, surveys and articles examine possible differences between stated preference measures of WTP and revealed preference measures, and among methods. Examples include \cite{KnetschDavis}, \\ \cite{CummingsBrookshire1986}, \cite{Wardman1988}, \cite{CummingsHarrison1995}, \\ \cite{ListGallet2001}, \cite{Hicks2002}, \cite{MurphyAllen2005}, \\ \cite{LuskSchroeder}, \cite{BernheimRangel2008}, \\ \cite{HorowitzMcConnell2008}, and \cite{dMKPZ2016}. In the somewhat different setting of consumer decisions about health, \cite{HepatitisB2015} compares stated and revealed preferences using a discrete choice experiment. 

\bigskip \noindent 
Survey methods vary. For instance, \cite{LuskSchroeder} compares WTP for better
treatment of animals raised for ham when the measurements were based on
choice experiments and on auction experiments. In another experiment consumers stated their bid on an item and
then, through coupons, could buy the product at different prices. Here \cite{LoureiroMcCluskey2003} found such 
measures in agreement with stated preferences for different types of apples. Other experiments which test for convergence of SWTP and RWTP include \cite{AdamowiczLouviere1994}, \cite{CummingsHarrison1995}, \cite{JohannessonLiljas1998}, \cite{BlumenscheinJohannesson1998}, \cite{OConorJohannesson1999}, \cite{Earnhart2002},\\  \cite{AzevedoHerringes2003}, \cite{CarsonMitchell2003}, \cite{BirolKontoleon2006}, and \\ \cite{Whitehead2006}.  

\bigskip \noindent 
In a comprehensive meta-analysis of studies that included both measures
based on stated preferences and measures based on revealed preferences, \\ \cite{CarsonFlores1996} found the ratio of revealed
WTP (RWTP) and stated WTP (SWTP) to average near 1. However this
ratio varied from near 0.5 to as high as 24 in different studies. The ratio also varied with the nature of the good, whether private or semi-public, and
the type of the good, whether an environmental amenity or some other
good, though there were only a few studies in some combinations of categories. 

\bigskip \noindent 
Reasons suggested for the divergence of stated measures of WTP from revealed WTP in the context of contingent valuation studies were first systematically reviewed in  \cite{MitchellCarson1989}.  
The mechanisms identified there remain the main concerns when eliciting and interpreting stated preferences.  In the discussion of our results, we will include agreement on the hypothetical setting, strategic behavior, categorical response, inexperience, and impersonal assessment.

\bigskip \noindent 
When a hypothetical setting is described, the respondent's understanding of the setting might differ from the one intended by the researcher who will later interpret the response.  It has been argued that experience with variables described in the hypothetical setting improves the consistency between the researcher's and respondent's interpretations. 

\bigskip \noindent 
When presented with a hypothetical setting, respondents may display strategic behavior by
either stating a preference as if it were a starting bid or by stating a preference
that is not a personal valuation in order to influence policy. 

\bigskip \noindent 
A categorical (or perhaps emotional) response is one that expresses a desire to either
conserve or use a resource regardless of the level of use or particular value suggested. 

\bigskip \noindent 
Inexperience may be either lack of familiarity with the setting for the
hypothetical question, lack of familiarity with or uncertainty regarding the consequences of a particular
choice, or a poor understanding of the survey itself -- see the discussions in \cite{Hanemann1999} and \cite{FischhoffFurby1988}. 

\bigskip \noindent 
An impersonal assessment
occurs when the consumer states a value that he or she believes to be fair,
in the sense of the fair market value or as it would be viewed by others, rather
than stating her or his own valuation.  See \\ \cite{PlottZeiler2005},
\cite{Chalmers2007}, and \cite{Carlsson2010} for related experiments and discussions. 

\bigskip \noindent 
When only SWTP is available, one would like to understand which of the various mechanisms that might cause a difference with RWTP is present.  Once the likely mechanisms are identified, one would want to understand how to adjust the measured SWTP to achieve a measure of WTP that can be used in a meaningful way for assessment and in forming policy.

\section{Stated and Revealed Measures In The RVF Study.}
\label{sec:experiment}

Our pilot study used responses to a questionnaire that had already been designed and was being given to visitors to a recreational site, the Rio Vista Falls park in San Marcos, Texas.  We were able to add a question to the survey that asked whether the consumer would continue travel to the site even if fuel prices were higher.  Thus we compare a revealed preference -- the cost calculated from actual travel, with a stated preference -- the statement whether the consumer was willingness to pay the additional travel cost associated with the hypothetical setting. 

The two measures offer a test of internal validity for the model of consumers' behavior used, and provide a test for the presence of various biases in the measure based on stated preference. 

Very generally, we examine, in the very specific setting of our survey, whether consumers can accept and understand hypothetical questions that would (most likely) lead to changes in their consumption, how consumers interpret the setting of the question, whether consumers are aware of their whole consumption bundle, and whether they truthfully state their WTP when questioned about a hypothetical situation?  We cannot answer any of these questions in isolation, but our model allows us to compare the aggregate behavior and to conclude that in this setting of costs associated with visits to recreational sites, consumer responses stating their preferences are consistent with their revealed preferences. 

\section{Revealed WTP From Travel.}
\label{sec:revealed}

During the summer of 2008, visitors to the San Marcos River, RVF and surrounding park areas were approached with an on-site survey questionnaire designed, and coordinated by Jenna McKnight Winters.  The survey is described later in the paper and asked about various aspects of interest to the City and others involved in managing the park and the river.  One question asked for the visitor's home zip code, allowing a calculation of the likelihood that a person residing a certain distance from RVF will travel that distance to visit.  

The group at a certain distance, the total population in this group as given in the 2000 census \cite{2000census}, and the number of visitors from this group in our sample are summarized in table \ref{table:popDist}. 

\begin{table}
\caption{Population and responses by distance}
\label{table:popDist}       
\begin{tabular}{lll}
\hline\noalign{\smallskip}
Distance (miles) & Population (2000) & Respondents  \\
\noalign{\smallskip}\hline\noalign{\smallskip}
10--50 & 3,435,123 & 142 \\
50--90 & none & none \\ 
90--130 & 1,594,679	 & 17 \\
130--170	& 4,095,308 &	16 \\
170--210 & 3,096,034 & 11 \\
210--250 & 4,918,126 & 6 \\
250--290 & 1,766,821 & 4 \\
\noalign{\smallskip}\hline
\end{tabular}
\end{table}

To build our model of travel costs, the remaining task is to convert the number of visitors into a probability that a resident, residing at the given distance from San Marcos, visits the RVF area.  To do this, we needed to estimate the number of visitors represented by each respondent.

As explained in the appendix on the survey and the use of the data, each respondent represented $454.4186$ annual visitors, and the probability of visiting was obtained by dividing the annual number of visitors by the number of residents in the corresponding distance group. 

Fitting the resulting data with an exponential function accounts for 87.734 percent of the variation and is statistically highly significant.  Statistically, we have an R value of 0.87734, and this correlation would occur at random, even if no relation existed, only 0.00589 of the time (based on the corresponding F statistic).  The probability that a resident spends $x$ dollars is 
\begin{equation}
{\rm Pr } = 0.0211457277071 \, e^{-0.0271286 \, x}\, .
\end{equation}
The price, $x$, was obtained from the distance traveled using the price of gas and fuel economy. The price of gas was approximately 4 dollars per gallon for the summer of 2008 in the region \cite{AAA2008} and the average fuel economy was 20.8 mpg \cite{EPA2008}.  (The distances summarized in table  \ref{table:popDist} are one-way distances obtained from \cite{distanceCalc}.)

The average RWTP obtained from this distribution is $0.77946254901$ dollars per person.  The total population within 290 miles of San Marcos was $18,906,091$ in 2000, which would give a total of about 14.74 million dollars over the lifetime of the population involved.  

Cesario \cite{Cesario1976} found that the value to individuals of the foregone travel time is about 20 percent of the travel expenses.  Using this multiplier gives (a lower bound for) the total RWTP in the sum of 17.68 million dollars.  This is an estimate of the contribution to RWTP from the value of visiting for the entire population within 290 miles of RVF.

\section{Stated WTP.}
\label{sec:stated}

The survey conducted in 2008 included a question as to whether visitors would reduce their number of visits should gas prices rise to 5 dollars a gallon.  We used the response to the question about additional expenses only for those visitors actually interviewed in RVF park itself, thinking that the reaction to the question might vary with location and hence be more accurate for those visitors.  We analyze the marginal costs that lead to this decrease in willingness to visit. 

Our estimate uses the responses from 167 visitors to RVF who traveled more than 10 miles and less than 290 miles, and who replied to the question asking whether they would visit less frequently should gas prices rise to 5 dollars (from the price at the time which was approximately 4 dollars per gallon).  

Of these 167 people, 55 said they would decrease their number of visits and 112 reported that they would accept the higher cost.  We calculated, using the distances traveled, that the average additional cost would be 3.985933 dollars.  With our data on the direct expenditure on gasoline $\Delta_{0} = 3.985933$, and with the multiplier accounting for the loss of time $\Delta = 4.7831196$.  

To calculate the probability that a tourist from a certain group, that is, from a group traveling a particular distance, agrees to pay the additional cost we simply divide the number of those stating that they will continue to travel by the total number of tourists from the group.  For the groups traveling $10$ to $50$ miles the probability was $p_1=82/121$, for those traveling $90$ to $130$ miles the probability was $p_2=10/14$, for those traveling $130$ to $170$ miles we found $p_3=9/14$,  for those traveling $170$ to $210$ miles $p_4=4/9$,  for those traveling $210$ to $250$ miles $p_5=4/5$, and for those traveling $250$ to $290$ miles the probability was $p_6=3/4$. 

Here equation (\ref{eq:Delta}) applies and so 
\begin{equation}
e^{-\lambda \Delta} = \Big[ \, \Big( \frac{82}{121} \Big)^{121}  \Big( \frac{10}{14} \Big)^{14}  \Big( \frac{9}{14} \Big)^{14}  \Big( \frac{4}{9} \Big)^{9}  \Big( \frac{4}{5} \Big)^{5}  \Big( \frac{3}{4} \Big)^{4} \, \Big] ^{1/167} = 0.6673615 . 
\end{equation}
Solving for the mean, $1/\lambda $, gives $11.9729166$ dollars per visitor.  

The amount above is an estimate of SWTP for each of the (estimated) 78,160 visitors to RVF, giving a total of 935,803.16 dollars.  This is an estimate of the annual contribution to SWTP from the value of visiting.

\section{Comparison of the Estimates.}
\label{sec:comparison}

To compare the estimate of RWTP for the population as a whole and the estimate for the annual value of SWTP, we need to adjust for changes in the population and for the duration corresponding to the estimates.

The population in the region near San Marcos has grown by about 2.5 percent each year between 2000 and 2007, according to American FactFinder \cite{2000census}, so the total RWTP would be multiplied by this increase (a factor of approximately 1.2184 over the 8 years since the 2000 census).  However, this means that the number of people in the population from which each visitor is self-selected is also larger, so the probability of being willing to travel the distance is decreased by the same factor.  In short, the estimate of RWTP reported above, 17.6 million, is relevant to the year of the survey, namely 2008 and needs no further adjustment.  

A rough estimate for the population's turnover is the time it takes for the number of new residents to equal the number of current residents.  The mortality rate is approximately 1.3 percent and the net growth rate for the region of interest here is 2.5 percent so the rate at which new residents join the population is about 3.8 percent.  Using this rate, the number of new residents is approximately the same as the number of current residents in $n$ years where $1.038^{n}=2$.  So the period is about 18.6 years. 

In conclusion, we estimated that RWTP for visiting the Rio Vista Falls area, including the section from City Park to Rio Vista Falls, is 17.6 million dollars in each 18.6 year period. 

The estimate of SWTP, obtained from the hypothetical scenario of travel despite a higher gas price, gave an annual value of 935,803 dollars.  Adjusting the annual value to a duration of 18.6 years (to which the estimate of RWTP applies) gives a total of approximately 17.41 million dollars for this period.  

The ratio of the estimate of SWTP to the estimate of RWTP is 17.41/17.68 or approximately 98.5 percent.  This suggests that the stated preferences and revealed preferences agree, that is, that respondents' reports on their behavior under hypothetical circumstances agree with their observed behavior.  

For the exponential distribution, the variance equals the mean, so for our sample (with an average of 17.41), a variation of 23.6 percent ($\sqrt{17.41} /17.68$) would correspond to one standard deviation.  Hence the difference obtained, of 1.5 percent, is about 0.065 of a standard deviation and agreement between the two estimates is statistically extremely good.

\section{Conclusions: Agreement of Parameters.}
\label{sec:agreement}

We used a probabilistic model of behavior to compare measures that are difficult to compare in a deterministic economic model, with the comparison achieved by comparing the statistical parameters that describe the behavior and that can easily be interpreted in monetary terms. In other words, as is more typically done for risk, we construct a probabilistic model of consumption, and we then use the revealed and stated observations to separately calibrate this model.  This contrasts with current practice which matches the data as a function rather than as a model. 

This modeling idea may serve well in answering other questions for which agreement among economic measures follows from agreement among parameters of a probabilistic model. 

Specifically, we used this approach to compare stated and revealed measures for WTP.  In this instance, the statistical parameters estimated using the stated and revealed data agree closely, which demonstrates that the mechanisms that could cause a divergence of the two measures, as discussed earlier,  cancelled one another, were  eliminated due to the design, or do not appear in this setting.  

In our study the value of a visit was clear to respondents.  Moreover, the value of a visit in the near future and the value of the present visit are likely similar.  The data agree with respondents finding the hypothetical situation credible and understanding the consequences to her or his behavior as requested by the questionnaire.  No incentives were put in place to correct for strategic responses, but policies regarding added costs or improvements were not directly at stake and we did not see an inflated value of SWTP.  Regarding a categorical response, a ``warm glow'' response to visiting the park and river would increase the proportion of respondents who state that they will continue to visit RVF even if the driving costs rise.  An increase in this proportion would also result from an endowment effect, as respondents would be reluctant to give up their current practice of visiting.  Since we found SWTP to be slightly lower than RWTP, the evidence suggests the absence of either a categorical response or an endowment effect. 

Finally effects in the survey may have cancelled.  For example, the respondents might have considered the reduction in their disposable income as a result of a rise in gas prices, therefore reducing SWTP, but an endowment effect might have raised SWTP, leading to a cancelation of the two effects.  The income effect is small, since expenditures on gasoline were approximately 4\% of household income, so such a combination leaves our result far from the factor of 2 or more typically found in studies showing increased SWTP. 

When conducting surveys about consumer behavior in some settings, it may well be possible to build a statistical model and use it to compare SWTP and RWTP.  Such studies could, in aggregate, improve methods for estimating WTP and, in particular, for estimating WTP when revealed preferences are inaccessible. 

Our method may also allow comparisons of other measures of social behavior by constructing a probabilistic statistical model of the underlying behavior and calibrating the model separately using the different measures. 

\section*{Appendix I: Project Description and Survey Details. }
\label{sec:surveys}

The San Marcos River begins in Aquarena Springs in the city of San Marcos, and continues for a few miles through the city.  The springs and a stretch of river between City Park and Rio Vista Falls Park are the two main attractions in San Marcos associated with the river.  Aquarena Springs is a reserve.  For recreation, including swimming, floating, and boating, the section between City Park and Rio Vista Falls is the principal venue.  Some ``tubing'', that is floating down river in inner tubes, continues down stream of RVF. 

The city of San Marcos renovated the Rio Vista dam in the spring of 2006 after significant flaws were found in the old dam during the winter of 2005-2006.  

\subsection*{The RVF renovation}
\label{sec:rvf}

The renovation included replacement of the old dam with a new one, repairs to the side structure of the dam on the river's right bank, and substantial changes in hard-scaping and landscaping of the river bed and banks for a stretch of about 300 feet below the dam. 

Renovations of the banks integrates the Rio Vista dam into two city parks nearby and provides easy access to the river on its right bank (access on the left bank near a picnic area and restaurant remains similar to what it was before the renovations). 

The renovation has turned the section of river immediately below the new dam into a white-water park which kayakers, inner tube swimmers, and canoeists can enjoy. 

The dam supports habitat for several species of fish, plants, and birds, including several endangered aquatic species.  The renovation is expected to continue this protection but also to allow some endangered species, such as the spring darter, easier passage into the protected habitat. 

\subsection*{Visitors' survey}
\label{sec:visitors}

During the summers of 2007 and 2008, visitors to the San Marcos River, RVF and surrounding park areas were approached with an on-site survey questionnaire.   Visitors whose trip originated from within and from outside of San Marcos were included in the survey.  

The questionnaire consisted of 31 questions.  Typical questions elicited the number of previous visits to the river, the length of stay, the number of people encountered, the number of members in the group, the purpose of the trip, total expenditures on the trip and accommodations, reactions to the density of visitors to the parks, reactions to the condition of the parks, and knowledge of natural resources in the park.  Those administering the survey recorded the time of the interview and its location. 

Several questions on the visitors survey were of particular interest in estimating WTP for travel.  One question asked for the visitor's home zip code. Others asked how many visitors were observed during the visit and the duration of the visit.  In addition, the survey provided data on whether or not the respondent had visited the Rio Vista Dam/Falls area before or since it had been rebuilt, and whether the visit included parks other than RVF.  In 2008 a question was added to the survey asking whether an increase in the cost of gasoline to five dollar per gallon would cause the visitor to visit less often. 

\subsection*{Data conversion}
\label{sec:dataDigest}

In estimating the number of visitors represented by each interview we confined our attention to the 172 respondents interviewed in RVF park itself and to responses approximating the number of people respondents encountered during the visit.  The interview date shows whether the visit occurred on a week day, a week-end day, or during a holiday.  

During holidays, some respondents reported seeing thousands of people.  We decided that such observations are unreliable (birders have counted thousands of birds in migrating flocks, after some training in observing flocks, but the respondents here were not particularly trained in estimating the size of a crowd).  Hence we conservatively confined our attention to the numbers reported on weekends and week days. 

Of the interviews, 68 were conducted during week days and a total of 8,034 visitors were reportedly seen.  During weekends, 51 respondents reported seeing 19,100 visitors.  For the 54 interviews conducted during holidays we used the same number of visitors as on weekends (a lower bound).  

Respondents reported an average stay of 4.029 hours.  Hence the week-day rate is estimated at (8034/68)/4.029 or approximately 29.3 visitors per hour and the weekend rate is estimated at approximately 93.0 visitors per hour. 

The active hours for visiting the RVF area are about 10 a.m. through 8 p.m. for a total of 10 hours per day.  From Memorial Day to Labor Day there were 29 weekend days and holidays and 75 week days, so the total number of visitors is estimated at 78,160 visitors per season. 

The Lions' Club has a facility renting inner tubes to people floating the San Marcos River from City Park to Rio Vista Falls and beyond (they are not the only rental source and some people bring their own tubes and boats). We were told by the manager of the Lions' Club inner tube rental outfit and by a City Parks official that this outfit has rented as many as 100,000 inner tubes in one season.  This makes our estimate of 78,160 visitors plausible, and, in fact, most likely very conservative. 

In calculating the likelihood that a person visits the RVF area given the distance of travel from her or his residence to San Marcos, we multiplied the number of respondents from a given distance group by $78,160/172 = 454.4186$ visitors per respondent and divided this number by the number of residents in this distance group. 

\section*{Appendix II: Justification for the Model}
\label{sec:modelJust}

While similar arguments to the one given below are common in the formulation of statistical models, we explain the assumption in this model.  In general terms, the assumption is that the likelihood of an additional expenditure is independent of the expenditure so far.  In concrete terms, a tourist who has traveled from point A to point B is now, having already spent the funds for this travel, as likely to continue to the tourism site as a person starting at point B.  Or said in an alternative way, the decision made by a person at point B is independent of how the person got to point B -- the travel process is memoryless. 

Suppose that a person has many portions to spend (we think of each portion as being very small).  In deciding the expenditure on a particular item the consumer is equally likely to spend any particular portion, until she or he decides to stop spending.  That is, the total expenditure is the number of portions spent on the item and the decision regarding each portion is independent of the decision regarding any other portion until a decision to stop spending is reached. 

Suppose, then, that the probability of spending any particular portion of (a small) size $m$ is $p$, with $0<p<1$ (and $p$ very near zero).  And suppose that the decision regarding each next portion is statistically independent.  Let $X$ denote the total amount spent.  Then the event $X \geq km$, that is the consumer decides to spend each of the first $k$ portions, has probability $p^k$.  Also $X= (k+1)m$ when the consumer spends the first $k+1$ portions and then decides against the $(k+2)^{\rm nd}$.  Thus $X=(k+1)m$ with probability $p^{k+1}(1-p)$.  The conditional probability is the ratio of these two.  Symbolically, 
\begin{equation}
{\rm Pr } \big( X = (k+1)m \, \big| \,   X \geq km \big)= p^{(k+1)} (1-p) \big/ p^k = p (1-p),
\end{equation}
and this probability is independent of $k$.  

The continuous distribution that describes the independence of $X \leq x + z$ from $X\leq x$ is the exponential distribution.  With mean $1/\lambda$ for a payment of $x$ the density function is $\lambda e^{-\lambda x}$, and the cumulative distribution function is $F(x)=1-e^{-\lambda x}$.  One can then calculate that 
\begin{equation}
\begin{tabular}{l} 
${\rm Pr } \big( x \leq X \leq x+z \, \big| \,   X \leq x \big)= \big( F(x+z)-F(x)\big) \big/  \big( 1 - F(x) \big) $ \\ 
\\ 
$ = \big( e^{-\lambda x}-e^{-\lambda (x+z)}\big) \big/ e^{-\lambda x} = 1 -e^{-\lambda z} = F(z).$
\end{tabular}
\end{equation}

The average value for this exponential distribution is $1/\lambda$ and this quantity is measured in units of expenditure per tourist.  This value, $1/\lambda$, can thus be easily estimated from data.  Furthermore, knowing $1/\lambda$ and the number of potential tourists gives a value for the total travel cost.  

\section*{Acknowledgements}
The authors thank Jenna McKnight Winters for allowing us to use her survey of visitors to the San Marcos River.


\begin{thebibliography}{99}

\bibitem[Adamowicz et~al.~(1994)]{AdamowiczLouviere1994}
Adamowicz, W., J. Louviere, and M. Williams. (1994). Combining revealed and stated preference methods for valuing environmental amenities. \textit{Journal of Environmental Economics and Management} \textbf{26}, 271--292.

\bibitem[Alberini and Chiabai (2007)]{AlberiniChiabai} 
Alberini, Anna, and Aline (2007). Urban environmental health and sensitive populations: How much are the Italians willing to pay to reduce their risks? \textit{Regional Science and Urban Economics}, \textbf{37}, 239--258. 

\bibitem[AAA (2008)]{AAA2008}
American~Automobile~Association (2008).  American Automobile Association report accessed during the summer of 2009 at www.fuelgaugereport.com/TXmetro.asp 

\bibitem[Azevedo et~al.~(2003)]{AzevedoHerringes2003}
Azevedo, C. D., J. A. Herriges, and C. I. Kling. (2003). Combining revealed and stated preferences: Consistency tests and their interpretations. textit{ American Journal of Agricultural Economics} \textbf{85}(3), 525--537.


\bibitem[Bernheim and Rangel (2008)]{BernheimRangel2008}	
Bernheim, B. D. and A. Rangel (2008). Beyond revealed preference: Choice theoretic foundations for behavioral welfare economics. \textit{NBER Working Paper Series}. National Bureau of Economic Research, Cambridge.


\bibitem[Bettman and Kakkar (1977)]{BettmanKakkar} 
Bettman, J.R., and Kakkar, P. (1977). Effects of information presentation format
on consumer information acquisition strategies. \textit{Journal of Consumer Research}, \textbf{3}, 233--240. 


\bibitem[Birol et~al.~(2006)]{BirolKontoleon2006} 
Birol, E., A. Kontoleon, M. Smale (2006). Combining revealed and stated preference methods to assess the private value of agrobiodiversity in hungarian home gardens. In \textit{Third World Congress of Environmental and Resource Economists}, 1--45. Kyoto, Japan. 


\bibitem[Bishop and Heberlein (1986)]{BishopHeberlein1986}
Bishop, R. C. and T. A. Heberlein (1986). Does contingent valuation work? valuing environmental goods: A state of the art assessment of the contingent valuation method. In Cummings, R. G., D. S. Brookshire, W. D. Schulze (eds.), \textit{Valuing Environmental Goods: An Assessment of the Contingent Valuation Method}.  Rowman and Allanheld, Totawa, NJ. 

\bibitem[Black (2018)]{Black2018}
Black, Katie J. (2018), Wide open spaces: Estimating the willingness to pay for adjacent preserved open space. \textit{Regional Science and Urban Economics}, \textbf{71}, 110--121. 
	
\bibitem[Blumenschein et~al.~(1998)]{BlumenscheinJohannesson1998}
Blumenschein, K., M. Johannesson, G. C. Blomquist, B. Liljas, and R. M. O'Conor. (1998). Experimental results on expressed certainty and hypothetical bias on contingent valuation. \textit{Southern Economic Journal} \textbf{65}(1), 169--77. 

\bibitem[Carlsson (2010)]{Carlsson2010}
Carlsson, F. (2010). Design of stated preference surveys: Is there more to learn from behavioral economics. \textit{Environmental Resource Economics}, \textbf{46}, 167--177. 

	
\bibitem[Carson et~al.~(1996)]{CarsonFlores1996}
Carson, R. T., N. E. Flores, K. M. Martin, and J. L. Wright. (1996). Contingent valuation and revealed preference methodologies: Comparing the estimates for quasi-public goods. \textit{Land Economics} \textbf{72}(1), 80--99. 

	
\bibitem[Carson et~al.~(2003)]{CarsonMitchell2003}
Carson, R. T., R. C. Mitchell, W. M. Hanemann, R. J. Kopp, S. Presser, and P. A. Ruud. (2003). Contingent valuation and lost passive use: Damages from the Exxon Valdez oil spill. \textit{Environmental and Resource Economics} \textbf{25}, 257--286. 


\bibitem[Cesario (1976)]{Cesario1976}
Cesario, F. J. (1976). Value of time in recreation benefit studies. \textit{Land Economics} \textbf{52} (1), 32--41. 


\bibitem[Chalmers (2007)]{Chalmers2007}
Chalmers, K. G. (2007). Can Information Narrow the Gap Between Stated and Revealed Preferences?  The Effect of Information on the Residential Location Process. \textit{Department of City and Regional Planning}. University of North Carolina, Chapel Hill. Masters: 44. 


\bibitem[Clawson (1959)]{Clawson1959}
Clawson, M.  (1959).  \textit{Methods of Measuring the Demand for and the Value of Outdoor Recreation}.  Reprint no. 10, Resources for the Future, Washington,  D.C. 


\bibitem[Clawson and Knetsch (1966)]{ClaK1966}
Clawson, M. and J. L. Knetsch (1966).  \textit{Economics of Outdoor Recreation}.  John Hopkins Press,  Baltimore, MD, Published for Resources for the Future 


\bibitem[Cummings et~al.~(1986)]{CummingsBrookshire1986}
Cummings, R. G., D. S. Brookshire, W. D. Schulze (eds.) (1986).  \textit{Valuing Environmental Goods: An Assessment of the Contingent Valuation Method}.  Rowman and Allanheld, Totawa, NJ. 

	
\bibitem[Cummings et~al.~(1995)]{CummingsHarrison1995}
Cummings, R. G., Harrison,  G. W. and E. E. Rutstr\"om. (1995). Homegrown values and hypothetical surveys: Is the dichotomous choice approach incentive-compatible? \textit{American Economic Review }, \textbf{85}(1): 260--266. 


\bibitem[DeMartino et~al.~(2016)]{dMKPZ2016}
De~Martino, S., F.~Kondylis, S.~Pagiola, and A.~Zwager (2016).  Do they do as they say? Stated versus revealed preferences and take up in an incentives for conservation program. \textit{PES Learning Paper 2015-1, Environment and Natural Resources}, World Bank, Washington, DC, USA. 


\bibitem[Distance Calculator (2009)]{distanceCalc}
Distance Calculator (2009).  We used http://www.infoplease.com/atlas/calculate-distance.html 


\bibitem[Earnhart (2002)]{Earnhart2002}
Earnhart, D. (2002). Combining revealed and stated data to examine housing decisions using discrete choice analysis. \textit{Journal of Urban Economics}, \textbf{51}(1), 143--169. 


\bibitem[EPA (2009)]{EPA2008}
EPA (2009). \textit{Light-Duty Automotive Technology and Fuel Economy Trends: 1975 Through 2008}.  Report by the Environmental Protection Agency, September 2008. Found at www.epa.gov/otaq/fetrends.htm 


\bibitem[Fischhoff and Furby (1988)]{FischhoffFurby1988}
Fischhoff, B. and Furby, L. (1988).  Measuring values: A conceptual framework for interpreting transactions with special reference to contingent valuation of visibility. \textit{Journal of Risk and Uncertainty}, \textbf{1}, 147--184. 


\bibitem[Hanemann (1999)]{Hanemann1999}
Hanemann, W.~M.~(1999).  The economic theory of WTP and WTA. In \textit{Valuing Environmental Preferences}, I.~J.~Bateman and K.~G.~Willis (eds.) Oxford University Press. 


\bibitem[Heberlein et~al.~(2005)]{HeberleinAnd2005}
Heberlein, T. A., M. A. Wilson, R. C. Bishop, and N. C. Schaeffer. (2005) 
Rethinking the scope test as a
criterion for validity in contingent valuation. \textit{Journal of
Environmental Economics and Management}, \textbf{50}, 1--22. 


\bibitem[Hicks (2002)]{Hicks2002}
Hicks, Robert L.~(2002). A comparison of stated and revealed preference methods for fisheries management. 
\textit{Journal American Agricultural Economics Association} (Since 2008: Agricultural and Applied Economics Association) 

	
\bibitem[Horowitz et~al.~(2008)]{HorowitzMcConnell2008}
Horowitz, J. K., K. E. McConnell, and J. J. Murphy. (2008). Behavioral foundations of environmental economics and valuation. In List, J.~and M.~Price (eds.), \textit{Handbook on Experimental Economics and the Environment},  1--45. Edward Elgar, Northampton, MA.

	
\bibitem[Johannesson et~al.~(1998)]{JohannessonLiljas1998}
Johannesson, M., B. Liljas, and P-O. Johansson. (1998). An experimental comparison of dichotomous choice contingent valuation questions and real purchase decisions. \textit{Applied Economics}, \textbf{30}(5), 643--647. 

\bibitem[Knetsch (1963)]{Knetsch1963}
Knetsch, J. L. (1963)  Outdoor recreation demand and values. \textit{Land Economics}, \textbf{39}, 387--396. 


\bibitem[Knetsch and Davis (1966)]{KnetschDavis}
Knetsch, J. L. and R. K. Davis (1966). \textit{Comparisons of Methods for Resource Evaluation. Water Research. A}. Kneese, V, and S. C. Smith. Johns Hopkins University Press, Baltimore, Resources for the Future. 

\bibitem[LaPiere (1934)]{LaPiere1934}
LaPiere, R. T. (1934). Attitudes vs. actions.'\textit{Social Forces}, \textbf{13}(2), 230--237. 

\bibitem[List and Gallet (2001)]{ListGallet2001}
List, J. A. and C. A. Gallet (2001). What experimental protocol influence disparities between actual and hypothetical stated values? \textit{Environmental and Resource Economics}, \textbf{20}, 241--254. 

\bibitem[Loureiro et~al.~(2003)]{LoureiroMcCluskey2003}
Loureiro, M. L., J. J. McCluskey, and R. C. Mittelhammer. (2003). Are stated preferences good predictors of market behavior? \textit{Land Economics}, \textbf{79}(1), 44--55.

\bibitem[Lusk and Schroeder (2006)]{LuskSchroeder}
Lusk, J.L., and T.C. Schroeder. (2006). Auction bids and shopping choices. \textit{Advances in
Economics Analysis and Policy}, \textbf{6}(1), Art.4. 

\bibitem[Mattijs et~al.~(2015)]{HepatitisB2015}
Mattijs S Lambooij, Irene A Harmsen, Jorien Veldwijk, Hester de Melker, Liesbeth Mollema, Yolanda WM van Weert, and G Ardine de Wit (2015). Consistency between stated and revealed
preferences: A discrete choice experiment and a
behavioural experiment on vaccination behaviour
compared. \textit{BMC Medical Research Methodology}, \textbf{5:19}.  DOI 10.1186/s12874-015-0010-5 

\bibitem[Mitchell and Carson (1989)]{MitchellCarson1989}
Mitchell, Robert Cameron, and Richard T.~Carson (1989). \textit{Using Surveys to Value Public Goods: The Contingent Valuation Method}.  Washington, D.C., Resources for the Future. 
	
\bibitem[Murphy et~al.~(2005)]{MurphyAllen2005}
Murphy, J. J., Allen, P. G., T. H. Stevens, and D. Weatherhead (2005). A meta-analysis of hypothetical bias in stated preference valuation. \textit{Environmental and Resource Economics}, \textbf{30}, 313--325. 

\bibitem[Nishitateno and Burke (2021)]{NishitatenoBurke} 
Nishitateno, Shuhei, and Paul J.~Burke (2021). Willingness to pay for clean air: Evidence from diesel vehicle registration restrictions in Japan. \textit{Regional Science and Urban Economics}, \textbf{88}, article 103657. 
	
\bibitem[O'Conor et~al.~(1999)]{OConorJohannesson1999}
O'Conor, R. M., M. Johannesson, and P-O. Johansson (1999). Stated preferences, real behaviour and anchoring: Some empirical evidence. \textit{Environmental and Resource Economics} \textbf{13}, 235--248.


\bibitem[Plott and Zeiler (2005)]{PlottZeiler2005}
Plott, C.~R.~and K. Zeiler  (2005). The willingness to pay-willingness to accept gap, the
`endowment effect,' subject misconceptions, and experimental procedures for eliciting valuations. \textit{The American Economic Review} \textbf{95}, 330--345.


\bibitem[Schuman and Johnson (1976)]{SchumanJohnson1976}
Schuman, H. and M. P. Johnson (1976). Attitudes and behavior. \textit{Annual Review of Sociology. A. Inkeles}. 
Palo Alto, Annual Reviews, Inc. \textbf{2}, 161--207.


\bibitem[US Census (2000)]{2000census}
U. S. Census Bureau (2008). American FactFinder,  The 2000 census data, and regional updates for 2007, were accessed during the summer of 2009 through American FactFinder at http://factfinder.census.gov 
 

\bibitem[Wardman (1988)]{Wardman1988}
Wardman, M. (1988). A comparison of revealed preference and stated preference models of travel behavior. \textit{Journal of Transport Economics and Policy}, \textbf{22}(1), 71--91. 

	
\bibitem[Whitehead (2006)]{Whitehead2006}
Whitehead, J. C. (2006). Improving willingness to pay estimates for quality improvements through joint estimation with quality perceptions. \textit{Southern Economic Journal} \textbf{73}(1), 100--111. 

\bibitem[Zhang and Zheng (2015)]{ZhangZheng2015} 
Zhang, Junfu, and Liang Zheng (2015). Are people willing to pay for less segregation? Evidence from U.S. internal migration. \textit{Regional Science and Urban Economics}, \textbf{53}, 97--112. 


\end{thebibliography}
\end{document}